\documentclass[journal]{IAENGtran}
\usepackage{bm, amsmath, amssymb}
\usepackage{color}
\usepackage{listings}    

\definecolor{dkgreen}{rgb}{0,0.6,0}
\definecolor{BlueDeFrance}{rgb}{0.19,0.55,0.91}
\definecolor{MyDarkBlue}{rgb}{0.0,0,0.7}
\definecolor{mygray}{rgb}{0.95,0.95,0.95}
\definecolor{brown}{rgb}{0.59,0.29,0}
\ifCLASSINFOpdf
   \usepackage[pdftex]{graphicx}
   \DeclareGraphicsExtensions{.pdf,.jpeg,.png}
\else
   \usepackage[dvips]{graphicx}
   \DeclareGraphicsExtensions{.eps}
\fi

\begin{document}
%
\title{A Python Class for Higher-Dimensional Schr{\"o}dinger Equations}
%
%
%

\author{Amna~Noreen,~\IAENGmembership{Member,~IAENG,}
	K{\aa}re Olaussen,~\IAENGmembership{Member,~IAENG,} 
\thanks{Manuscript received February 10, 2015. 
	}
\thanks{Amna Noreen is with the Division of
	Science and Technology, University of Education Township Campus, Lahore,
PK, Pakistan. e-mail: aamnanoreen12@gmail.com}
\thanks{K{\aa}re Olaussen is with the Department of Physics,
	Norwegian University of Science and Technology, Trondheim, N-7491, Norway. e-mail: Kare.Olaussen@ntnu.no}
}

\maketitle

\pagestyle{empty}
\thispagestyle{empty}

\begin{abstract}

We announce a Python class for numerical solution of
Schr{\"o}dinger equations in one or more space dimensions,
employing some recently developed general classes for numerical
solution of partial differential equations, and
routines from \texttt{numpy} and \texttt{scipy.sparse.linalg} (or 
\texttt{scipy.linalg} for smaller problems).

\end{abstract}

\begin{IAENGkeywords}
Quantum Mechanics, Anharmonic Oscillators, Sparse Scipy routines.
\end{IAENGkeywords}

%
\IAENGpeerreviewmaketitle

\section{Introduction}

\lstset{
	emphstyle=\color{MyDarkBlue}\bfseries,
	emph={shape,dim,size,bC,geometry,r0,rE,dr,set_bC,
		domain,targetNsource,narr,rvec,qarr,kvec,def_f,def_F,def_g,def_G,
		evalf,evalF,evalg,evalG,evalfn,evalFr,evalgq,evalGk,
		evalFr,values,fftvalues,FFT,iFFT,shift,restrict,prolong,
		lattice,matrix,linOp,varOp,stensOp,laplace,stensil,def_V}
}

%
%
%
%
\IAENGPARstart{S}{urprisingly} many basic problems from physics and other
natural sciences can be solved by one-dimensional analysis, mainly
due to the symmetries of nature. A successful approach is often to
search for symmetric solutions, or (through separation of variables
for linear problems) solutions which can be composed
of single-variable functions with simple symmetry behavior.


However, not all problems of interest enjoy a high degree of
symmetry. Many of them are still amenable to numerical analysis,
although not always in any easily available manner. In this note we
present an attempt to improve this situation for a particular
class of problems from Quantum Mechanics, eigenvalue equations
like
\begin{equation}
	\left[-\bm{\Delta} + V(\bm{r})\right] \psi_n(\bm{r})
	= E_n \psi_n(\bm{r}),
	\label{SchrodingerEquation}
\end{equation}
or variants and extensions of such equations.

Our own interests in this class of problems arose
when we attempted to generalize
our method of very-high-precision solutions of
such equations in one dimension
\cite{AsifAmnaKareIngjald, AmnaKareI,
AmnaKareII} to higher dimensions. Our method
works faster and most straightforward with some prior knowledge
of eigenvalues and eigenfunctions. For one-dimensional problems
such knowledge can f.i.~be obtained by use of the WKB method
\cite{AmnaKareIII, AmnaKareIV}. The corresponding information
is less accessible through analytic means in higher dimensions
\cite{BanksBenderWu, BanksBender}. A numerical approach looks
like a faster and more general strategy, which for most
practical applications may anyway be sufficient. 

We believe that a numerical solver
where the user can simply provide
an essentially analytic formula for $V(\bm{r})$ in \eqref{SchrodingerEquation},
plus some information related boundary conditions and desired accuracy
of the numerical approximation, would
be of general interest.

This can be realized using the freely
available packages in \texttt{\bfseries numpy} \cite{Walt_etal} and
\texttt{\bfseries scipy} \cite{Jones_etal, Oliphant}. Three
Python classes developed to simplify the
numerical solution of partial differentials
in general have recently been
announced \cite{Mushtaq_etal}.
We have refactored our original code to make use of these classes.

For many problems of interesting complexity
and size, the resulting code can be run on
a normal laptop -- and of course on more powerful
computer systems.

\section{Basic numerical implementation}

Our code is constructed as a Python class, \lstinline!SchrodingerEquation!
which inherits the \lstinline!LatticeOperator! class, and requires
the \lstinline!Lattice! and \lstinline!LatticeFunction! classes,
plus routines from the \texttt{\bfseries numpy} and \texttt{\bfseries scipy} packages.
The latter packages are freely available for Windows, Mac OSX, and
Linux platforms, or in (perhaps) numerically more efficient
commercial versions. They allow a relatively simple formulation
and organization of the problem in Python, with all heavy
numerical calculations delegated to fast compiled routines.

\subsection{Lattice shape and geometry}

For discretization the continuum of positions $\bm{r}$
is replaced by a rectangular $d$-dimensional lattice of 
\lstinline!shape!
\begin{equation*}
   \bm{N} = (N_0,...,N_{d-1}),
\end{equation*}
so that the total number of lattice points is
$N= \prod_{k=0}^{d-1} N_k$.
It is our experience that a modern high-end laptop can
handle models for which $N \le \;\text{\emph{few}} \times 10^8$.
This allows resonably accurate treatment of one-particle problems in $\le 4$
dimensions, two-particle problems in $\le 2$ dimensions,
and three-particle problems in one dimension.

Note that in Python all array positions are counted from zero.
Hence each point of this lattice is described by an
index vector
\begin{equation*}
	\bm{n} = (n_0,\ldots,n_{d-1}),
	\quad\text{where}\quad
	0 \le n_k < N_k.
\end{equation*}

Geometry can be introduced through a mapping
$\bm{n} \rightarrow \bm{r}_{\bm{n}}$.
The definition of a lattice model, along with
a simple set of possible (rectangular) geometries,
is implemented in the \lstinline!Lattice! class.
Two \lstinline!dim!-dimensional parameters,
\lstinline!rE! and \lstinline!r0!,
can be given. They specify respectively
the length of each lattice side (by default 1),
and the position of the ``lower left'' corner
of space (by default the origin). Note that all
lattice points lie \emph{inside} the spatial
region defined in this way, half a lattice
cell away from the edges.

\subsection{Boundary conditions}

Since the lattice must necessarily be finite
for any finite-memory computer, it is necessary
to define boundary conditions \lstinline!bC! on $\psi$ at its edges,
or more precisely how $\psi$ should be extended
beyond the edges.
Some choices commonly used in Quantum
Mechanics are: 
\begin{enumerate}
	\item Periodic extension, \lstinline!'P'!.
	\item Symmetric extension, \lstinline!'S'!.
	\item Antisymmetric extension, \lstinline!'A'!.
	\item Extension with zero values, \lstinline!'Z'!.
\end{enumerate}
These choices may be different in different directions of
the lattice, and --- except for \lstinline!'P'! --- different
at the two edges of a given direction.
All the above boundary conditions are implemented in
the \lstinline!Lattice! class.

\subsection{Lattice Laplacian. Stensil representations.}

The simplest implementation of a lattice Laplacian
in one dimension is by use of the common
discretization formula
\begin{equation}
	\psi''(x_n) \approx
	\frac{\psi(x_{n+1}) -2 \psi(x_n) - \psi(x_{n-1})}{\Delta_x^2},
\end{equation}
where $\Delta_x$ is the distance between nearest-neighbor lattice
points. The generalizes to the standard $(2d+1)$-stensil definition
of the lattice Laplacian,
\begin{align}
  &(\Delta_L \psi)(\bm{r}_{\bm{n}}) =\nonumber\\ 
  &\sum_{k=0}^{d-1} \left[
  {\psi(\bm{r}_{\bm{n}_{(+k)}}) - 2\psi(\bm{r}_{\bm{n}})
  + \psi(\bm{r}_{\bm{n}_{(-k)}}})\right]/{dr_k^2}.
  \label{2Dplus1stensil}
\end{align}
Here
\(
	dr_k = \text{\lstinline!dr[k]!}
\)
is the sidelength of the lattice cell in $k^{\text{th}}$ direction, and
\begin{align*} 
    \bm{n}_{(+k)} &\equiv (n_0,\cdots,n_{k}+1,\cdots,n_{d-1}),\\
    \bm{n}_{(-k)} &\equiv (n_0,\cdots,n_{k}-1,\cdots,n_{d-1}).
\end{align*}
The discretization error of \eqref{2Dplus1stensil} is of order 
\begin{equation}
  \varepsilon_{\Delta} = \mathcal{O}(\,\sum_{k=0}^{d-1} dr^2_k\,).
\end{equation}
With periodic boundary conditions equation~\eqref{2Dplus1stensil}
is straightforward to implement in \texttt{\bfseries numpy} by
use of the \lstinline!roll! operation.
Other boundary conditions require more
considerations and code, which have been incorporated
in the \lstinline!Lattice! and \lstinline!LatticeOperator!
classes. An arbitrary (short-range) operator $O$, represented by
a stensil $s_O(\bm{b})$ such that
\begin{equation}
    \left(O \psi\right)(\bm{r}_{\bm{n}}) = 
    \sum_{\bm{b}} s_O(\bm{b})\, \psi(\bm{r}_{\bm{n}-\bm{b}}),
\end{equation}
is supported.
Here $\psi(\bm{r}_{\bm{n}-\bm{b}})$ is interpreted according
to its boundary conditions when $\bm{n}-\bm{b}$ falls outside
the lattice. The stensil \lstinline!sO! can be any $d$-dimensional
\texttt{\bfseries numpy} array provided by the user (but the program will run slowly
if it is too large, since the sum over $\bm{b}$ is executed
in native Python). By default \lstinline!sO! is set to the stensil
defined by equation~\eqref{2Dplus1stensil}.

\section{Simple implementation}

It follows from the discussion above that
a basic implementation of the \lstinline!SchrodingerEquation!
class is quite straightforward. The lattice parameters
can be set up by use of the \lstinline!Lattice! class, and a
standard approximation of the Laplace operator in
equation~\eqref{SchrodingerEquation} is already available
in the \lstinline!LatticeOperator! class (or an alternative
sensil can be provided).

We further need to define an array representing the potential
$V(\bm{r}_{n})$ on every site of the lattice; this can be
done in various ways by methods in the \lstinline!LatticeFunction! class.

We combine these to a
method \lstinline!TplusVstensil(psi)! in the class
\lstinline!SchrodingerEquation!. This performs the mapping
\begin{equation}
   \psi(\bm{r}_{\bm{n}}) \rightarrow 
 -(\Delta_L\psi)(\bm{r}_{\bm{n}})
	+ V(\bm{r}_{\bm{n}})\,\psi(\bm{r}_{\bm{n}})
\end{equation}
for any given input array $\psi(\bm{r}_{\bm{n}})$.
This method can be used (i) to construct an explicit
matrix representation of the linear operator in question,
using the \lstinline!matrix()! method, for
further analysis by standard dense matrix routines
in \texttt{\bfseries scipy.linalg},
or (ii) as a \lstinline!LinearOperator! alternative to an explicit sparse
matrix, for further analysis by the appropriate iterative
routines in \texttt{\bfseries scipy.sparse.linalg}.
The implementation of the metod is straightforward and short:

\vspace{-4ex}

\begin{lstlisting}
def TplusVstensil(self, psi):
	"""Return (T + V) psi."""
	psiO = -self.stensOp(psi)
	if self.def_V is not None:
		psiO += self.valuesV*psi
	return psiO
\end{lstlisting}
The bulk of the code lies in the method \lstinline!stensOp!, inherited from
\lstinline!LatticeOperator!, and methods in \lstinline!Lattice!
called by this routine. Note that all lattice dimensions and sizes
(restricted by available computer memory), and all combinations of
boundary conditions discussed above (ten possibilities for each
direction), is handled transparently to the user. The stensil
used to represent the operator $T$ is stored in the \lstinline!dim!-dimensional
array \lstinline!self.stensil!. It is by default the standard $(2d+1)$-stensil
for the lattice Laplacian, but is easily changed to any preferred array.

\subsection{Example: One-dimensional harmonic oscillator}

Consider the eigenvalue problem of the one-dimensional
harmonic oscillator,
\begin{equation}
    -\psi_n''(x) + x^2\,\psi_n(x) = E_n\,\psi_n(x).
    \label{HarmonicOscillator}
\end{equation}
The eigenvalues are $E_n = 2n+1$ for $n = 0, 1,\ldots$,
and the extent of the wavefunction $\psi_n(x)$ can
be estimated from the requirement that a classical
particle of energy $E_n$ is restricted
to $x^2 \le E_n$. A quantum particle requires a
little more space. 

The numerical analysis of this problem is
indicated by the code snippet 

\vspace{-4ex}

\begin{lstlisting}
myL = Lattice(shape=(2**7,), bC='Z', rE=(25, ), r0=(-12.5, ))
defV = lambda x: x[0]**2
myS = SchrodingerEquation(myL, def_V=defV)
myS.varOp = myS.TplusVstensil
H = myS.matrix()
evals = eigvalsh(H)
\end{lstlisting}
In lines 1 we define a geometric region of size $25$,
with origin in the middle, and cover it with
a one-dimensional lattice of $2^7=128$ sites.
In line 2 we define the potential $V=x^2$, and use
this and \lstinline!myL! to define
an \emph{instance} of \lstinline!SchrodingerEquation!  
in lines 3. The property \lstinline!varOp! specifies
a operator used by \lstinline!SchrodingerEquation! in
many situations. This is set to \lstinline!TplusVstensil!
(different from the default) in line 4.
A $2^7 \times 2^7$ dense matrix representation, 
\lstinline!H!, of the operator is calculated
in line 5. Finally all eigenvalues of \lstinline!H!
are computed in line 6. Here almost all computational work
is executed in lines 5--6.

\begin{figure}[h!]
\begin{center}
\includegraphics{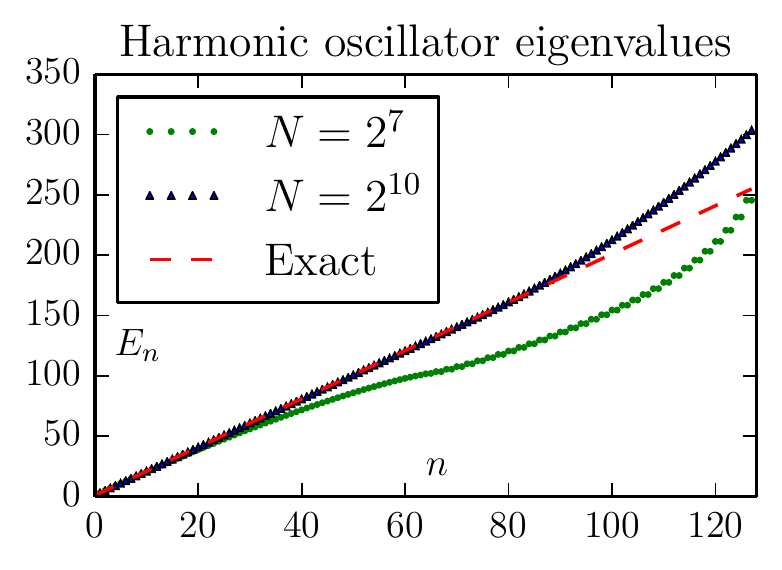}
\end{center}
\caption{\label{figure1}
The 128 lowest eigenvalues of equation~\eqref{HarmonicOscillator},
computed with the standard $3$-stensil approximation for the Laplace
operator (here the kinetic energy $T$). The parameters are chosen
to illustrate two typical
effects: With \lstinline!bC='Z'! boundary conditions the harmonic
oscillator potential is effectively changed to $V=\infty$ for
$x\ge 12.5$, thereby modifying the behavior of extended (highly
exited) states. The effect of this is to increase the eigenenergies
of such states, to a behavior more similar to a particle-in-box.
This is visible for $n \gtrsim 80$.
The effect of using the $3$-stensil approximation for $T$ is
to change the spectrum of this operator from $k^2$ to (the slower
rising) $(2/\Delta_x)^2\sin^2 (k\Delta_x/2)$. This is visible
in the sublinear rise of the spectrum for $N=2^7$.
}
\end{figure}

The behavior of the resulting eigenvalues is shown in Fig.~\ref{figure1}.

\begin{figure}[h!]
\begin{center}
	\includegraphics{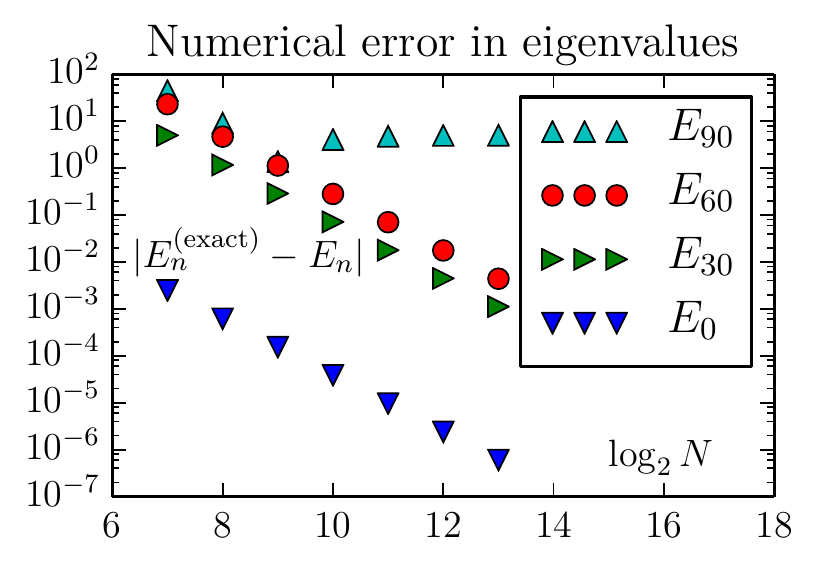}
\end{center}
\caption{\label{figure2}
The discretization error of energy eigenvalues
when using the standard $3$-stensil approximation for
the one-dimensional Laplace operator (here the kinetic energy
$E$). There is no improvement in $E_{90}$ beyond
certain $N$, because the corresponding oscillator
state is too large for the geometric region.
For the other states the improvement is consistent
with the expectation of an error proportional to
$\Delta_x^2$. This predicts an accuracy improvement
of magnitude $2^{12}=4096$ when the number of
lattice sites increases from $N=2^7$ to $N=2^{13}$
for a fixed geometry. The eigenvalues are computed
by the dense matrix routine \lstinline!eigvalsh!
from \texttt{\bfseries scipy.linalg}.
}
\end{figure}

For a better quantitative assessment we plot
some energy differences,
$E^{(\text{exact})}_n - E_n$, in Fig.~\ref{figure2}.

This \emph{brute force} method leads to
a dramatic increase in memory requirement with increasing
lattice size.
For a lattice with $N=2^m$ sites, the matrix
requires storage of $4^m$ double precision (8 byte)
numbers. For $m=13$ this corresponds to about
$\frac{1}{2}\;\text{Gb}$ of memory, for $m=14$
about $2\;\text{Gb}$. The situation becomes
even worse in higher dimensions.

Assuming that we are only interested some of
the lowest eigenvalues, an alternative approach
is to calculate these by iterative
routine \lstinline!eigsh! from 
\texttt{\bfseries scipy.sparse.linalg}.
The most straightforward change is to
replace lines 5--6 in the previous code
with the snippet

\begin{figure}[!h]
\begin{center}
	\includegraphics{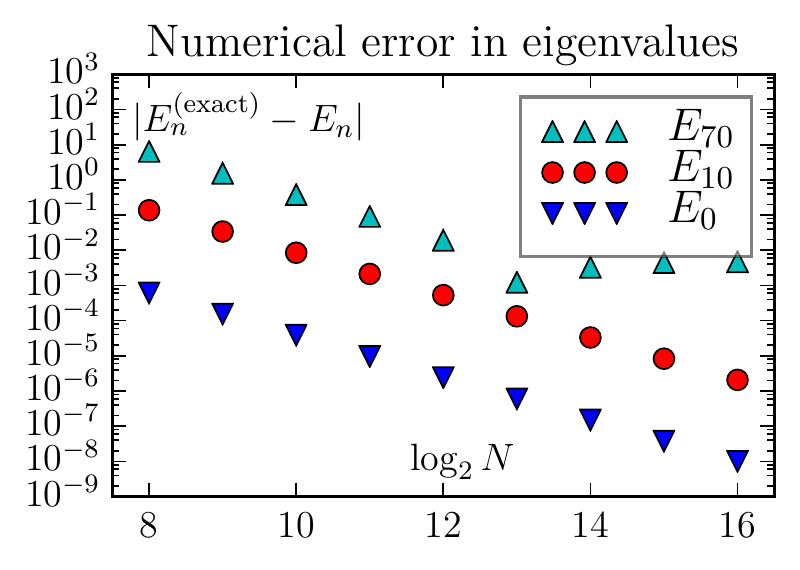}
\end{center}
\caption{\label{figure3}
The discretization error computed by the
routine \lstinline!eigsh! from
\texttt{\bfseries scipy.sparse.linalg}.
For a fixed lattice size the discretization
error is essentially the same as with dense
matrix routines. However, with a memory requirement
proportional to the lattice size (instead of its
square) it becomes possible to go to much larger
lattices. This figure also demonstrates ($E_{70}$)
that the error can be limited by boundary effects
instead of a finite discretization length $\Delta_x$.
}
\end{figure}

\vspace{-4ex}

\begin{lstlisting}[firstnumber=5]
N = myL.size
H = LinearOperator((N,N), matvec=myS.linOp, dtype=float)
evals = eigsh(H,which='SM',tol=10**(-8), k=128, return_eigenvectors=False)
\end{lstlisting} 
which will compute the lowest 128 eigenvalues
(still a rather generous amount). This allows
extension to larger lattices, as shown in Fig.~\ref{figure3}.

\begin{figure}[h!]
\begin{center}
	\includegraphics{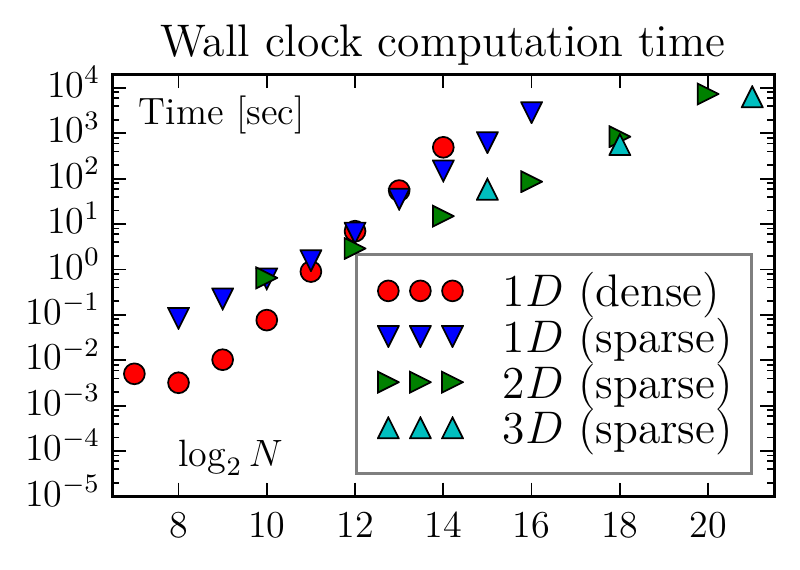}
\end{center}
\caption{\label{figure4}
The wall clock time used to find the
lowest 128 eigenvalues, for various
systems and methods. We have also used
the dense matrix routine \lstinline!eigvalsh! to compute the
eigenvalues of a $2^7 \times 2^7$ ($N = 2^{14}$)
two-dimensional lattice; not unexpected it takes
the same time as for a $2^{14}$ one-dimensional lattice.
Somewhat surprisingly, with \lstinline!eigsh! it is much faster to
find the eigenvalues for two-dimensional lattice than for
a one-dimensional with the same number of sites, and
somewhat faster to find the eigenvalues for a three-dimensional
lattice than for the two-dimensional with the same number
of sites.
}
\end{figure}

With a sparse eigenvalue solver the calculation
becomes limited by available
computation time, which often is a much weaker
constraint. With proper planning and organization
of calculations the relevant timescale is the
time to analyze and publish results (i.e.~weeks
or months). The computation time is nevertheless
of interest (it shouldn't be
years \cite{MyrheimOlaussen}). We have measured the wall clock time
used to perform the computations for
Figs.~\ref{figure2}--\ref{figure3}, performed
on a 2012 Mac Mini with 16 Gb of memory, and
equipped with a parallellized
\texttt{\bfseries scipy} library. Hence, the
\lstinline!eigvalsh! and \lstinline!eigsh!
routines are running with 4 threads.
The results are plotted in Fig.~\ref{figure4}.

\begin{figure}[h!]
\begin{center}
	\includegraphics{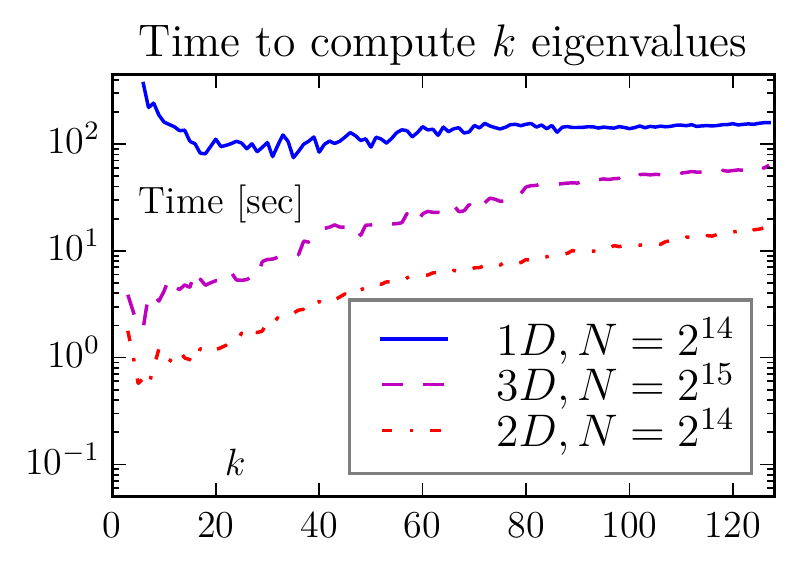}
\end{center}
\caption{\label{figure5}
One may think that it takes longer
to compute more eigenvalues. This is
not always the case when the number
of eigenvalues is small, as demonstrated
by this figure. The default choice of 
\lstinline!eigsh! is to compute $k=6$
eigenvalues. For our two- and three-dimensional
problems this looks close to the optimal value,
but it is too low for the one-dimensional problem.
}
\end{figure}

Here we have used the \lstinline!eigsh! routine in the most straightforward
manner, using default settings for most parameters. This means,
in particular, that the initial vector
for the iteration (and the subsequent set of trial vectors) may
not be chosen in a optimal manner for our category of problems.
It is interesting to observe that \lstinline!eigsh! works better
for higher-dimensional problems. The (brief) \texttt{\bfseries scipy}
documentation \cite{eigsh_documentation} says that the underlying
routines works best when computing eigenvalues of largest magnitude,
which are of no physical interest for our type of problems. It is
our experience that the suggested strategy, of using
the \emph{shift-invert} mode instead, does not work right out-of-the-box
for problems of interesting size (i.e., where dense solvers
cannot be used). We were surprised to observe that the computation
time may \emph{decrease} if the number of
computed eigenvalues increases, cf.~Fig.~\ref{figure5}.

\subsection{Example: 2- and 3-dimensional harmonic oscillators}

The $d$-dimensional harmonic oscillator
\begin{equation}
	\left[ -\boldmath{\Delta} + \bm{r}^2\right] \psi_n(\bm{r}) = 
	E_n\,\psi_n(\bm{r}),
\end{equation}
has eigenvalues $E_n = (d + 2n)$, for $n=0, 1, \ldots$. The degeneracy of
the energy level $E_n$ is $g_n = (n+1)$ in two dimensions,
and $g_n = \frac{1}{2}(n+1)(n+2)$ in three dimensions\footnote{The
general formula is $g_n = \binom{d-1+n}{d-1}$.}. These degeneracies
may be significantly broken by the numerical approximation.
For a numerical solution we only have to change lines 1--2 of  the previous
code snippet to

\vspace{-3.5ex}
\begin{lstlisting}
myL = Lattice(shape=(2**7,)*2, bC='Z', rE=(25, )*2, r0=(-12.5, )*2)
defV = lambda x: x[0]**2 + x[1]**2
\end{lstlisting}
in two dimensions, and

\vspace{-3.5ex}
\begin{lstlisting}
myL = Lattice(shape=(2**7,)*3, bC='Z', rE=(25, )*3, r0=(-12.5, )*3)
defV = lambda x: x[0]**2 + x[1]**2 + x[2]**2
\end{lstlisting}
in three dimensions.

\begin{figure}[h!]
\begin{center}
	\includegraphics{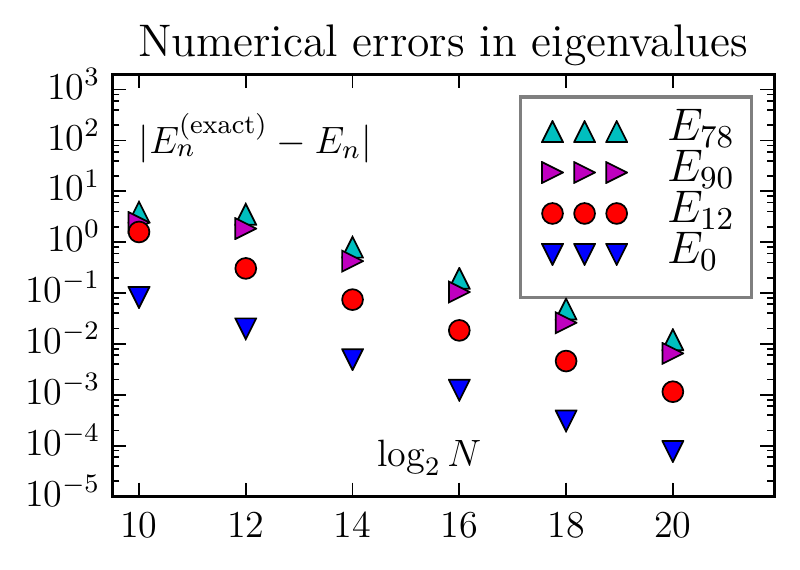}
\end{center}
\caption{\label{figure6}
The discretization error of energy eigenvalues
when using the standard 5-stensil approximation
for the two-dimensional Laplace operator.
Exactly, the states $E_{78}$ and $E_{90}$ are the two edges
of a $13$-member multiplet with energy $26$, and the
state $E_{12}$ is the middle member of a $5$-member multiplet with
energy $10$. With the chosen parameters all states considered
a well confined inside the geometric region; hence we do not
observe any boundary correction effects. 
}
\end{figure}

\begin{figure}[h!]
\begin{center}
	\includegraphics{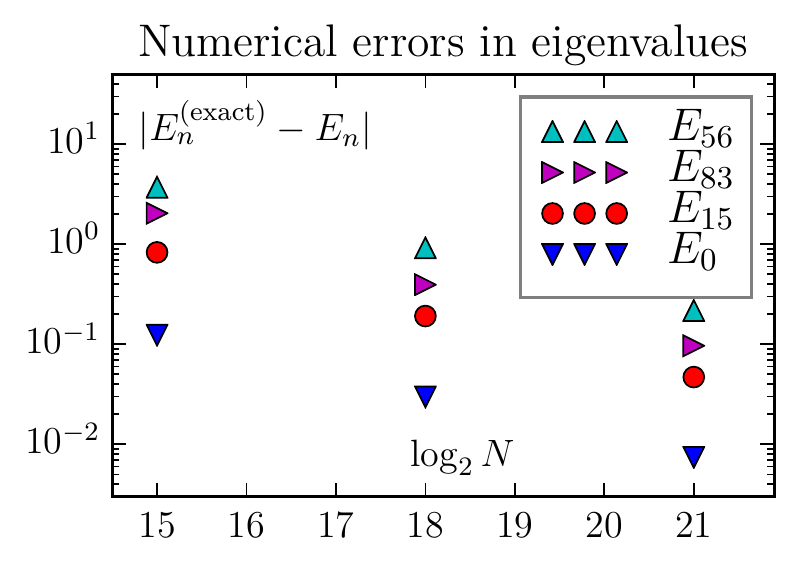}
\end{center}
\caption{\label{figure7}
The discretization error of energy eigenvalues
when using the standard 7-stensil approximation
for the three-dimensional Laplace operator.
Exactly, the states $E_{56}$ and $E_{83}$ are the two edges
of a $28$-member multiplet with energy $15$, and the
state $E_{15}$ is the middle member of a $10$-member multiplet with
energy $9$. 
}
\end{figure}

As already discussed, the routine 
\lstinline!eigsh! works somewhat faster in higher
dimensions than in one dimension (for the same total
number $N$ of lattice points). The corresponding
discretization error is shown in Figs.~\ref{figure6}-\ref{figure7},

The discretization error continues to scale
like $\Delta_x^2$. This means that a reduction of this error
by a factor $2^2=4$ requires an increase in the number of
lattice points by a factor $2^d$ in $d$ dimensions. This means that is
becomes more urgent to use a better representation of
the Laplace operator in higher dimensions.
Fortunately, as we shall see in the next sections, better
representations are available for our type of problems.

\section{FFT calculation of the Laplace operator}

One improvement is to use the reflection symmetry
of each axis ($x\to -x$, $y\to -y$, etc.) to reduce
the size of the spatial domain. This reduces $\Delta_x$ by a half,
without changing the number of lattice points.

A much more dramatic improvement is to use some variant of
a fast fourier transform (fft):
After a Fourier transformation,
$\psi(\bm{r}) \rightarrow \tilde{\psi}(\bm{k})$,
the Laplace operator turns into multiplication,
$
   \left(-\Delta \psi\right)(\bm{r}) \rightarrow 
   \bm{k}^2\,\tilde{\psi}(\bm{k})
$.
This means that application of the Laplace operator
can be represented by (i) a Fourier transform, followed
by (ii) multiplication by $\bm{k}^2$, and finally
(iii) an inverse Fourier transform. Essentially the
same procedure works for the related trigonometric
transforms.

These are also practical procedures for lattice
approximations, due to the existence of efficient
and accurate\footnote{The error of a back-and-forth
FFT is a few times the numerical accuracy, i.e.~in
the range $10^{-14}-10^{-15}$ with double precision numbers.
However, when an error of this order is multiplied
by $\bm{k}^2$ it can be amplified by several orders
of magnitude. Hence, the range of $\bm{k}^2$-values
should not be chosen significantly larger than required
to represent $\psi(\bm{r})$ to sufficient accuracy.
}
algoritms for discrete fourier and
trigonometric transforms. The time to perform
the above procedure is not significantly longer
than the corresponding stensil operations. The benefit
is that the Laplace operator becomes exact on the space
of functions which can be represented by
the modes included in the discrete transform.

\begin{figure}[h!]
\begin{center}
	\includegraphics{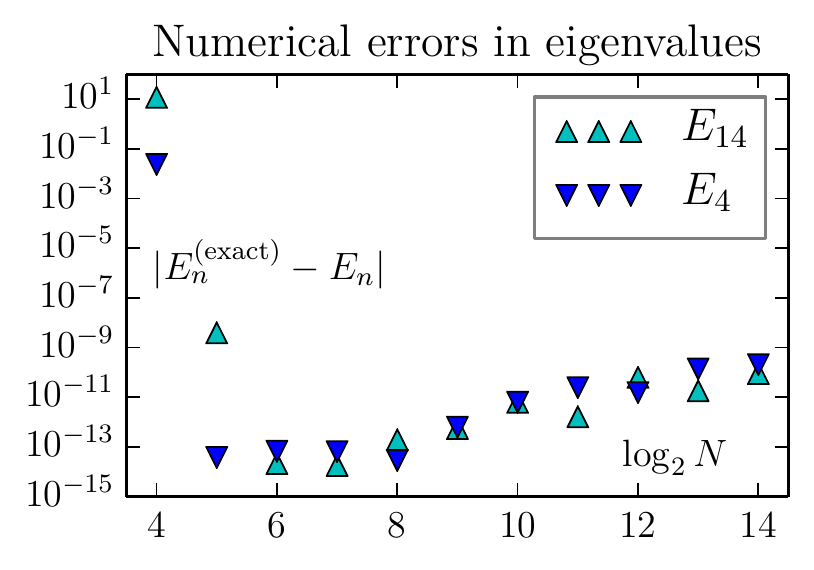}
\end{center}
\caption{\label{figure8}
With a FFT representation of the Laplace operator
the discretization error drops exceptionally fast
with $\Delta_x \propto N^{-1}$. When it becomes ``small enough''
the effect of numerical roundoff becomes visible;
the latter leads to an \emph{increase} in error
with $\Delta_x$. The results in this figure is for a one-dimensional
lattice, but the behavior is the same in all dimensions. The lesson
is that we should make $\Delta_x$ ``small enough'' (which in general
may be difficult to determine \emph{a priori}), but not smaller. It
may be possible to rewrite the eigenvalue problem to a form with
less amplification of roundoff errors.
}
\end{figure}

\begin{figure}[h!]
\begin{center}
	\includegraphics{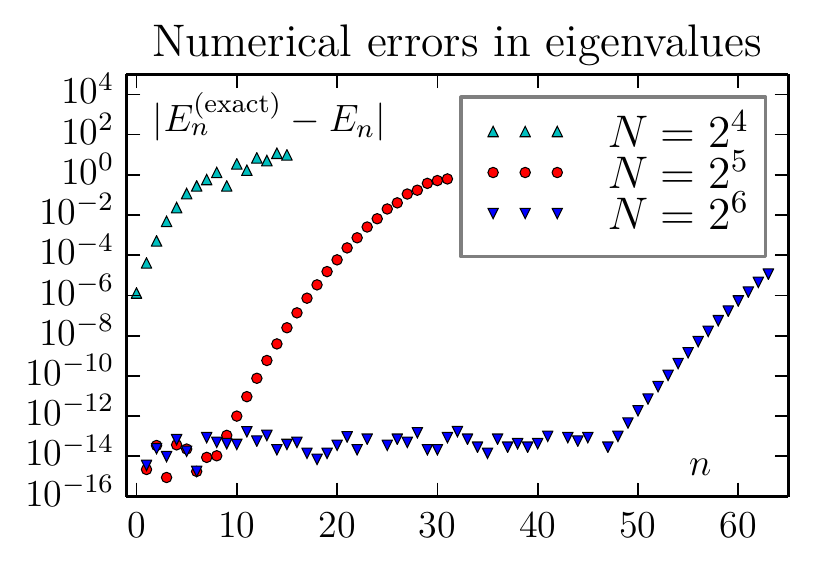}
\end{center}
\caption{\label{figure9}
Accuracy of computed eigenvalues for a $1D$ oscillator.
This figure may suggest that an increase in the number
of lattice size $N$ will lead to a accurate treatment of
states with higher $n$. Our findings are that
this is \emph{not} the case: The results for $N=2^7$ and
$N=2^8$ have essentially the same behavior as for $N=2^6$.
}
\end{figure}

\begin{figure}[h!]
\begin{center}
		\includegraphics{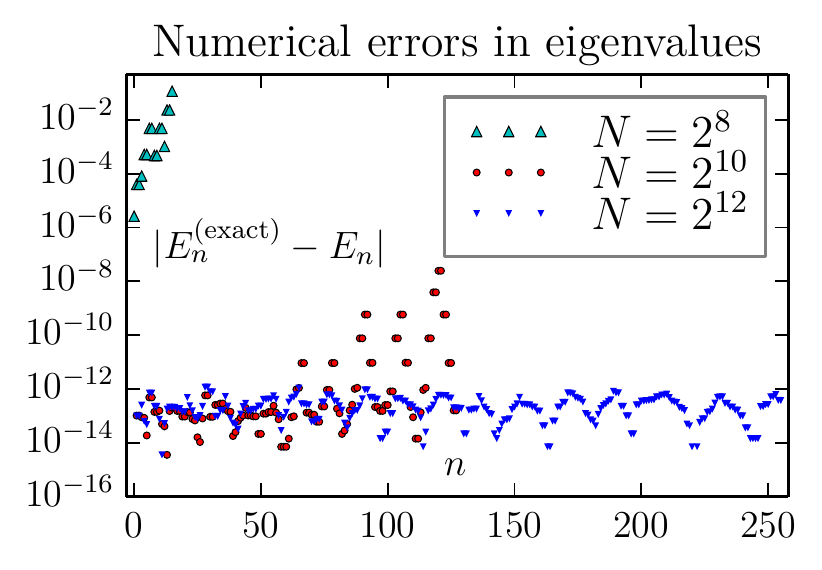}
\end{center}
\caption{\label{figure10}
Accuracy of computed eigenvalues for a $2D$ oscillator.
As can be seen, a large number of the lowest eigenvalues
can be computed to an absolute accuracy in the
range $10^{-14}$--$10^{-12}$ with lattice of
size $2^{6} \times 2^{6}$. We observe not improvement
by going to $2^{7} \times 2^{7}$ lattice, but no harm
either (except for an increase in the wall clock execution time
from about 3 to 30 seconds for each combination of boundary
conditions).
}
\end{figure}

\begin{figure}[h!]
\begin{center}
	\includegraphics{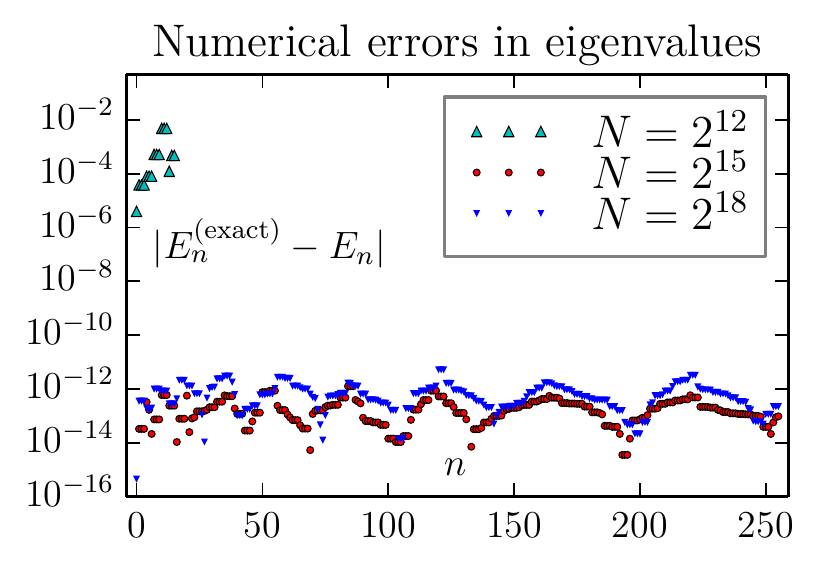}
\end{center}
\caption{\label{figure11}
Accuracy of computed eigenvalues for a $3D$ oscillator.
As can be seen, a large number of the lowest eigenvalues
can be computed to an absolute accuracy in the
range $10^{-14}$--$10^{-12}$ with lattice of
size $2^{6} \times 2^{6} \times 2^{6}$. We observe no improvement
by going to $2^{7} \times 2^{7} \times 2^{7}$ lattice, but no harm
either (except for an increase in the wall clock execution time
from about 6 to 95 minutes for each combination of boundary
conditions).}
\end{figure}

With the \lstinline!SchrodingerEquation! class it is easy
to employ the FFT representation.
This is built into the method \lstinline!TplusV!,
which is the default setting for \lstinline!varOp!.
Thus we only have to comment out (or remove) line 4 in
the first code snippet above:

\vspace{-4ex}

\begin{lstlisting}[firstnumber=4]
# myS.varOp = myS.TplusVstensil
\end{lstlisting}

The obtainable accuracy with this procedure increases
dramatically, as illustrated in Figs.~\ref{figure8}-\ref{figure11}.

It might be that the harmonic oscillator systems are particulary
favorable for application of the FFT representation.
One important feature is that the fourier components of the
harmonic oscillator wavefunctions vanishes exponentially
fast, like $\text{e}^{-\bm{k}^2/2}$, with increasing
wavenumbers $\bm{k}^2$. This feature is shared with all
eigenfunctions of polynomial potential Schr{\"o}dinger
equations, but usually with different powers of $\bm{k}$ in
the exponent (which may lead to a quantitative different behavior).

Further, for systems with singular wavefunctions the corresponding
fourier components may vanish only algebraically with $\bm{k}^2$.
The dramatic increase in accuracy cannot be expected for such cases. 

\section{Anharmonic oscillators} 

The \lstinline!SchrodingerEquation! class works with any computable
potential. All one has to do is to change the definition
of the function assigned to \lstinline!def_V!. However, in most
cases we no longer know the exact answer; this makes it difficult
to assess the accuracy of the result. 

One simple test is to consider two-dimensional potentials
of the form
\begin{equation}
	V(x,y) = x^4 + c\,x^2 y^2 + y^4,
\end{equation}
for various values of $c$. For $c=0$ and $c=6$ this system
an be separated to a set of two one-dimensional anharmonic
oscillators (with $x^4$ type potential), and for $c=2$
it can be separated in cylinder coordinates. Here comparison
with essentially exact solutions of the separated one-dimensional
problems can be made. Even in the absence of these, one may check
for degeneracies in the spectrum.

\section{Boundary conditions for radial operators}

For some problems one may perform a partial (or full)
symmetry reduction of equation~\eqref{SchrodingerEquation}. An example is
\begin{align}
   &\left[-\left(\frac{d^2}{dr^2} +
   \frac{1}{r}\frac{d}{dr} +
   \frac{d^2}{dz^2}\right) + \frac{\ell^2}{r^2} + V(r,z)\right]\, \psi(r,z)
   \nonumber\\
   &\qquad\qquad\qquad\qquad\qquad\qquad\qquad = E\,\psi(r,z).
   \label{RadialEquation}
\end{align}
In this case there is a natural boundary at $r=0$. This is
also a singular line for the equation.

What is the natural boundary condition at $r=0$?
Equation \eqref{RadialEquation} is often symmetric
under $r \to -r$, depending on the form of $V$.
With this symmetry equation~\eqref{RadialEquation} may be extended
to $r < 0$, and the solutions classified according
to their transformation under $r \to -r$. This
makes symmetric or antisymmetric boundary
conditions the natural choice.

Surprisingly, discussions of numerical approximation schemes
for (very common) singular equations like \eqref{RadialEquation} are
difficult to find in the textbook literature.
It may be tempting to introduce a new wavefunction,
$\psi(r,z) = r^{-1/2}\,\varphi(r,z)$, in order to
eliminate the first order derivative in \eqref{RadialEquation}.
This would transform the equation into the general form
\eqref{SchrodingerEquation}. However, this would also make
the searched-for solution $\varphi(r, z)$ singular at $r=0$,
hence difficult to approximate numerically.

\section{Domains of general shape}

There is also a convenient way to define domains
of general shape in \texttt{\bfseries numpy}, by specifying a
boolan vector which is \lstinline!True! for all
points in the domain, and \lstinline!False! for
all points outside. In this case we find only
the \lstinline!'Z'! boundary conditions to be a general and
unambiguous option. Stensil approximations of the Laplace
operator may be the best choice for such cases.

\section*{Acknowledgements}
We thank Dr.~Asif Mushtaq for useful discussions.
We also acknowledge support provided by Statoil
via Roger Sollie, through a professor II grant
in Applied mathematical physics.






%
%
%
%
\ifCLASSOPTIONcaptionsoff
  \newpage
\fi



%

\end{document}